\documentclass{aa}
\usepackage{graphicx}
\usepackage{txfonts}
\usepackage{soul}
\usepackage{dcolumn}
\defcitealias{Dea14}{D14}

\begin{document}

\title{An alternative stable solution for the
Kepler-419 system, obtained with the use of a genetic algorithm}

\titlerunning{A stable solution for the
Kepler-419 system}

\author{D. D. Carpintero
        \inst{1}  
        \and 
        M. Melita
        \inst{2}}

\institute{Instituto de Astrof\'isica de La Plata,
CONICET--UNLP, La Plata, Argentina \\
Facultad de Ciencias Astron\'omicas y Geof\'isicas, Univ.
Nacional de La Plata, La Plata, Argentina\\
\email{ddc@fcaglp.unlp.edu.ar}
\and
Instituto de Astronom\'ia y F\'isica del Espacio, CONICET--UBA,
Buenos Aires, Argentina \\
Facultad de Ciencias Astron\'omicas y Geof\'isicas, Univ.
Nacional de La
Plata, La Plata, Argentina\\
\email{melita@iafe.uba.ar} 
}

\date{Received Y,YY; accepted X,XX}

\abstract{The mid-transit times of an exoplanet may be non-periodic. The
variations in the timing of the transits with respect to a single period, that
is, the transit timing variations (TTVs), can sometimes be attributed to
perturbations by other exoplanets present in the system, which may or may not
transit the star.}
{Our aim is to compute the mass and the six orbital elements of an non-transiting
exoplanet, given only the central times of transit of the transiting body. We also aim
to recover the mass of the star and the mass and orbital elements of the
transiting exoplanet, suitably modified in order to decrease
the deviation between the observed and the computed transit times by  as much as possible.}
{We have applied our method, based on a genetic algorithm, to the Kepler-419
system.} 
{We were able to compute all fourteen free parameters of the system, which, when
integrated in time, give transits within the
observational errors. We 
also studied the dynamics and the long-term orbital evolution of the Kepler-419
planetary system as defined by the orbital elements computed by us, in order to
determine its stability.}{}


\maketitle

\section{Introduction}

The vast majority of the extra-solar planets known to date were
discovered by the method of transits, that is, by detecting the variation of the
luminosity of a star due to the eclipse produced when a planet crosses
the line of sight. A major contributor of these discoveries was
the {\sl Kepler} space mission, which monitored about $170\,000$ stars in
search of planetary companions \citep{Borucki2016,Coughlinetal2016}. The
detection of variations in the timing of the transits with respect to a mean
period (transit time variations, or TTVs), can be attributed to perturbations by
other planetary bodies in the system \citep{ASSC05,HM05}.

In order to compute the mass and orbital elements of the perturbing bodies from
the TTVs, an inverse problem must be solved; at present, only a handful of
extrasolar planets have been detected and characterized by solving this
inverse problem. Fast inversion methods were developed and tested by
\citet{NesvornyMorbidelli2008} and \citet{NesvornyandBeauge2010} with excellent
results for planets in moderately eccentric orbits. \citet{N09} also
extended the algorithm to the case of eccentric and inclined orbits. These
methods can be briefly summarized as follows. A deviation function is defined as
the difference between the modeled transit times and the observed ones; subsequently,
the downhill simplex method \citep[e.g.,][]{PTVF92} is used to search for its
minima. There can be a lack of generality if the modeled mid-transit times are
computed by means of a particular truncated planetary perturbation theory to
speed up calculations, for example when a particular multibody mean motion
resonance occurs, as can be the case of co-orbitals \citep[see for
example][]{NesvornyMorbidelli2008}. Also, as an alternative to the fast
inversion methods, a search for the elements of the perturbing body can also be
made by direct $N$-body integration, subsequently refining the result with an
optimization algorithm \citep{SA05,Becker2015,AS07}. In particular,
\citet{Betal14} solved the inverse problem for the Kepler-9 and Kepler-11
systems by applying a variety of techniques, including a genetic algorithm,
though none of the reported results were obtained with the latter. In each
experiment, they adjusted a subset of the parameters, keeping the remaining ones fixed. It
is worth mentioning that their best solution was obtained by excluding the
radial velocities, which suggests that the large errors of these data may
contribute to spoiling the fit. A common feature of most methods \citep[see
also][]{Carpinteroetal2014} is that they search in a seven-parameter space for the
mass and orbital elements of the unseen planet, whereas they fix the values of
the orbital elements and the masses of the transiting planet and of the star,
which are estimated before any search is done.

At present, as said, the number of exoplanets discovered by the TTV method is
small. The \url{exoplanet.eu} database, up to July 1$^{\rm }$ 2018, lists
only seven exoplanets discovered solely based on
the analysis of the TTVs time 
series. The first discovery was that of Kepler-46c, with an orbital period of
57.0 d, an eccentricity of 0.0145 and a mass of 0.37 M$_{\rm J}$, where M$_{\rm
J}$ is the mass of Jupiter \citep{Nesvornyetal2012}. In the case of Kepler-51
\citep{Masuda2014}, two transiting planets were already known and their 
TTV 
series revealed the existence of a third planetary body, KOI-602.02, with a mass
of 0.024 M$_{\rm J}$, an eccentricity of 0.008, and an orbital period of 130.19
d; later it was discovered that KOI-602.02 is also a transiting body. All planetary
components in this system have low orbital eccentricities. In the case of the
WASP-47 system, there are three transiting planets \citep{Adamsetal2015}, two of
them exhibiting measurable TTVs due to the existence of an additional body. To
estimate the mass and orbital properties of WASP-47e, the fourth planet in the
system, \citet{Becker2015} first modeled the TTV series of WASP-47b and
WASP-47d using {\sc ttvfast} \citep{DAHN14} and then used a Markov Chain
Monte Carlo algorithm \citep{FHLG13} to search for the solution of the inverse
problem. This methodology allowed them to put an upper limit on the mass of
WASP-47e. It is worth noting that \citet{Becker2015} used the TTVs not only to
determine orbital elements, but also to determine masses, therefore allowing the
confirmation of the planetary nature of the transiting objects. Finally, the
case of the Kepler-419 system is the most peculiar one, due to the large
eccentricity of the transiting planet, with a reported value of 0.833, a mass of
2.5 M$_{\rm J}$ , and a period of 69.76 d \citep{Dea12}. Its TTV time series
revealed the existence of a more distant and massive planet, with an approximate
orbital period of 675.5 days \citep[][hereafter \citetalias{Dea14}]{Dea14}.

The peculiarities of the presently known exoplanetary systems in general, and in
multiple systems in particular, place interesting constraints on the formation
scenarios that may produce the observed distributions of various parameters as
semimajor axes, eccentricities, masses, planetary radii, and so on. In particular,
the formation and evolution scenario of high-eccentricity planets is, at
present, quite heavily debated in the literature \citep{JuricTremaine2008,
FordRasio2008, Matsumuraetal2008, Moeckeletal2008, MalmbergDavies2009,
TerquemAjmia2010, Idaetal2013, TeyssandierTerquem2014, DuffellChiang2015}.
Naturally, the refinement in the knowledge of the masses and orbital elements of
the known exoplanets would allow more realistic investigations to be performed,
for example regarding their stability \citep[see for
example][]{TN14,MK16,MCB16}. 

It was recognized early on that the case of the Kepler-419 system offers a superb
model to solve the inverse problem \citep{Dea12}. Firstly, the interaction
between the two planets is secular in nature. This can be corroborated
through their orbital evolution since the semi-major axes remain constant and
the eccentricity librates with a single constant period, which guarantees that
the problem is not degenerate. If the interaction contained mean-motion
resonant terms, which depend on the mean longitudes of the system, different
combinations of these fast angles might result in the same characteristic
period, and therefore the solution would not be unique. Secondly, the time span
of the data covers about two periods of the more distant, non-transiting planet,
so enough information is available to solve the inverse problem accurately.

In order to solve this inverse problem, we use a genetic algorithm \citep{C95}
and seek maxima of a so-called fitness function, defined as the reciprocal
of the sum of the squares of the deviations between the modeled transit times
and the observed ones. The efficiency of the method is such that the problem can
be solved very accurately even if the number of unknowns is increased to include
all the dynamical parameters of the system, that is, the orbital elements and
masses of the two planets and the mass of the star, with the exception of the
longitude of the ascending node of the transiting planet, which can be fixed
arbitrarily. To obtain the modeled mid-transit times used to compute the
fitness function, we integrate the three-body gravitational problem in an
inertial frame. We disregard interactions due to body tides, relativistic
approximations, and non-gravitational terms arising from radiation transfer
between the star and the planets. 

By using as initial guesses sets of parameters close to those previously
estimated for the Kepler-419 planetary system, we have been able to 
find a solution that, when integrated in time, gives the
observed central transit times entirely within the observational errors. The
solution consists in a set of orbital parameters of both planets and their
masses and the mass of the central star.
We then
studied the long-term orbital evolution of the Kepler-419 planetary system as
defined by the orbital elements computed by us, thus characterizing its general
dynamics and determining its stability.

\section{Method}

\subsection{The genetic algorithm}

Since the space of parameters is multidimensional, an
optimization algorithm based on random searches is inescapable. We refrained
from using the popular MCMC algorithm because it requires some prior information
about the distribution of the parameters, which in our case is completely
unknown and probably very far from a simple multidimensional Gaussian. Instead,
we chose to work with a genetic algorithm approach that does not require any
previous knowledge of the background distribution. 
A genetic algorithm is an optimization technique that incorporates, in a
mathematical language, the notion of biological evolution. One of its remarkable
features is its ability to avoid getting stuck in local maxima, a characteristic
which is very important in solving our problem (see Sect. \ref{exres}). We used a genetic algorithm
in this work based on {\sc pikaia} \citep{C95}, which we
briefly describe here.

As a first step, the optimization problem has to be coded as a \emph{fitness}
function, that is a function $f:D\to\mathbb{R}$, where the domain $D$ is the
multidimensional space of the $n$ unknown parameters of the problem, and such
that $f$ has a maximum at the point corresponding to the optimal values. Let
${\bf x}$ be a point in this domain, which is called an \emph{individual}. The
algorithm starts by disseminating $K$ individuals ${\bf x}_i$, $i=1,\dots, K$ at
random. The set of $K$ individuals is called the \emph{population}, and, at this
stage, they represent the first \emph{generation}. Two members of the
population are then chosen to be \emph{parents} by selecting them at random but with
probabilities that depend on their fitnesses. The coordinates of the parents are
subjected to mathematical operations resembling the crossover of genes and mutation.
The resulting two points correspond to two new individuals (i.e., two new points
${\bf x}_i \in D$). Subsequently, a new pair of parents are chosen, not necessarily
different from earlier parents, and the cycle is repeated until a number $K$ of
offspring, that is, a new generation, have been generated. The new individuals will be,
on
average,  fitter than those of the first generation \citep{C95}. The loop
starts again from the selection of a pair of parents, and the procedure
continues until a preset number of generations has passed, or until a preset
tolerance in the value of the maximum of the fitness function is achieved. The
fittest individual of the last generation constitutes the result, which in
general will not be an exact answer, but an approximation to it. For a detailed
account of the numerical procedures, we refer the reader to \cite{C95}.

\subsection{Setup and computation of the mid-transit times}

The problem is stated as follows: given the (non-periodic) observed transit
times of a planet $t_{{\rm obs},i},\;i=1,\dots,N$, and assuming the presence of
a second, unseen planet which is held responsible for the lack of periodicity,
we want to find the mass and the six orbital parameters of each planet, and the mass of the
central star.

\begin{figure}
 \includegraphics[width=\columnwidth]{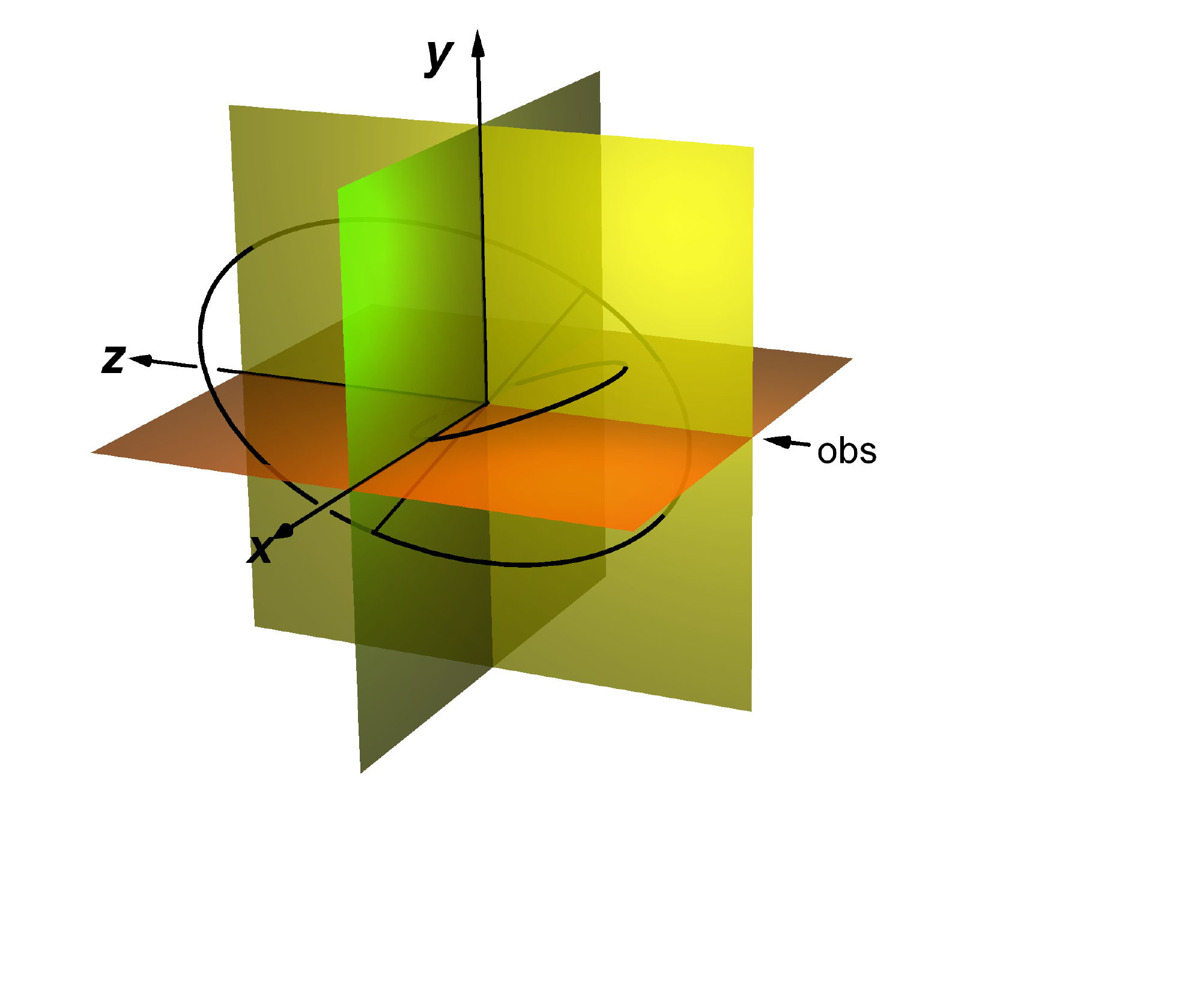}
 \caption{Geometry of the problem. The plane of the sky is the plane of
reference of the orbits, which we choose to be the $(x,y)$ plane. The observer
looks towards the positive $z$ axis, the center of mass of the system being at the origin of
coordinates. A transiting planet (small ellipse) has an inclination close to
$90^\circ$, that is, it has an orbit almost perpendicular to the plane of the sky.
Another planet (big ellipse) perturbs it. The ascending node of the transiting
planet defines the $x$ axis on the plane of the sky; the line of the nodes of
the other planet is marked with a black segment on this plane.}
 \label{fig1}
\end{figure}

The geometry of the problem is set as follows. Using Cartesian coordinates, we
take the plane of the sky as the $(x,y)$ plane (the reference plane); the $z$
axis points from the observer to the sky. On the reference plane, we set the $x$
axis as pointing to the direction of the ascending node of the transiting planet
(TP), that is, towards the point at which it crosses the plane of the sky moving
away from the observer. We note that this defines the longitude of the ascending
node of this planet as zero. The $y$ axis is chosen so that a right-handed basis
is defined (Fig.~\ref{fig1}). The perturbing planet has no constraints.

Let ${\bf p}=\{a,e,i,\Omega,\omega,M,m\}$ be the semimajor axis,
eccentricity, inclination, longitude of the ascending node, argument of the
periastron, mean anomaly and mass of a planet, respectively, which we simply refer to as
``elements'' for brevity. Given the elements of the TP ${\bf p}_{\rm t}$, those
of the perturbing planet ${\bf p}_{\rm p}$, and the mass of the star $m_\star$,
one can integrate the respective orbits, and compute the transits $t_{{\rm
com},i},\;i=1,\dots,N$ of the TP. The fitness function $F[{\bf t}_{\rm com}({\bf
p}_{\rm t},{\bf p}_{\rm p},m_\star),{\bf t}_{\rm obs}]$ is defined in our
problem as

\begin{equation}
F=\left[\sum_{i=1}^N 
(t_{{\rm obs},i}-t_{{\rm com},i})^2
\right]^{-1}.
\label{fitness}
\end{equation}
The larger the value of $F$, the better the solution. A value $F\to\infty$
would indicate that a set $\{{\bf p}_{\rm t},{\bf p}_{\rm
p},m_\star\}$ had been chosen so that the resulting transits
would coincide perfectly with the (central
values of) the observed ones; this would be the "exact" solution. In terms of the genetic algorithm, an individual ${\bf
x}$ is a set $\{{\bf p}_{\rm t},{\bf p}_{\rm p},m_\star\}$.

Once the first generation is generated at random, each individual is taken in
turn and its fitness is computed by integrating the respective orbits. To this
end, we first transform the elements of the planets to Cartesian coordinates,
with the center of mass at the origin
\citep[e.g.][]{MD99}. In this inertial frame and 
from these initial conditions we integrate the system as a full three-body
problem using the standard equations of motion where for each body $i$ (TP,
perturbing planet and star),
\begin{equation} 
\ddot{\bf r}_i=
-G m_j \frac{{\bf r}_j - {\bf r}_i}{|{\bf r}_j - {\bf r}_i|^3} 
-G m_k \frac{{\bf r}_k - {\bf r}_i}{|{\bf r}_k - {\bf r}_i|^3},
\end{equation}
with $G$ the gravitational constant, and $m_{j,k},\, {\bf r}_{j,k}$ the masses
and position vectors of the other two bodies. These allow an easy computation of
the transits, that is, the times when the TP is in the $z<0$ semispace and the $x$
coordinate of the TP, $x_{\rm t}$, and that of the star, $x_\star$, coincide.
We note that the star moves during the integration, so the transits are not, in
general, the instants when $x_{\rm t}=0$. To determine the instants of transit,
we used H\'enon's method of landing exactly on a given plane (in our case, the
plane $x=x_\star$) in only one backstep after the plane was crossed, a method
that he developed to compute surfaces of section \citep[][see Appendix
\ref{henon}]{H82}. The numerical integrations were carried out using a
Bulirsh-Stoer integrator \citep[]{PTVF92} with a variable time
step initially set at $\Delta t=0.005\, \rm{yr}\simeq$ 1.83 d; the relative
energy conservation was always below $10^{-11}$. 

\begin{figure}
 \includegraphics[width=\columnwidth]{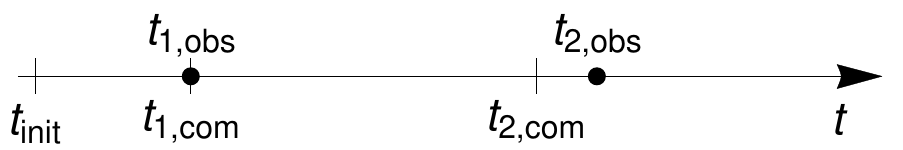}
 \caption{Time axis for each individual. The initial time of the integration and
  the first computed transits are marked with vertical lines (lower labels); the
  observed transits are marked with dots (upper labels). At the first computed
  transit, a Julian date is assigned to $t_{1,{\rm com}}$ corresponding to
  $t_{1,{\rm obs}}$. In this way, the instant $t_{\rm init}$ defining the
  elements gets a BJD$_{\rm TDB}$ equal to the first transit minus the time
  elapsed since the start of the integration.} 
 \label{tiempo}
\end{figure}

Each integration starts wherever the orbital elements put the planets ($t_{\rm 
init}$, see Fig. \ref{tiempo}), that is, the elements corresponding to each
individual are defined at $t_{\rm init}$. So far, this time has no Julian date
assigned. When the first transit $t_{1,{\rm com}}$ is found during the
integration, this instant is made to coincide with the first observed transit
$t_{1,{\rm obs}}$, therefore receiving the BJD$_{\rm TDB}$ 2\,454\,959.3308
label, that is, the date of the first observed transit reported in
\citetalias{Dea14}. In this way, the $t_{\rm init}$ instant also gets a
BJD$_{\rm TDB}$, thus defining the date to which the elements belong. Later, the
integration comes to a second transit $t_{2,{\rm com}}$, which would not be in
general coincident to the observed $t_{2,{\rm obs}}$ -- unless the initial
elements are the exact ones. The integration continues until $N$ transits have
been computed (where $N$ coincides with the number of observed transits); then,
the fitness of the individual is computed by means of Eq.~(\ref{fitness}). 

We briefly mention here the values of the parameters of the genetic algorithm
used in our experiments \citep[for details of their meaning, we refer the reader
to][]{C95}. We chose $K=5000$ individuals per generation, six (decimal) digits
to define the genotype of each parameter, a probability of genetic crossover
equal to 0.95, a generational replacement in which, after a new full generation
has been computed, the $K$ best individuals among the new and the old
generations are taken, and we have selected the parents with a probability
directly proportional to their fitnesses. We also chose a selection pressure of
80\%\ of the maximum possible value\footnote{The selection
pressure 
controls how much the fitness influences the probability of being selected as a
new parent.}. 
The mutation rate was variable, from
0.5 to 5\%. The algorithm was considered finished when
2000 generations
had passed.

The entire algorithm, written in {\sc fortran}, was parallelized with the {\sc
mpi} paradigm. Using four Intel Core i7 processors at 2.30 GHz, a typical run
of 2000 generations lasts approximately
14 hours.

\section{Experiments and results}
\label{exres}

We applied our algorithm to the system Kepler-419 \citepalias{Dea14}. This
system has at least two planets: a transiting warm Jupiter (Kepler-419b) and a
nontransiting super-Jupiter (Kepler-419c). The transiting planet has a period
of approximately 70 days, and there are $N=21$ observed transits, spanning the
first 16 quarters of the Kepler data.

We generated the initial conditions for each of our individuals by randomly
choosing values from a given interval for each element. The intervals were
initially chosen centered on the values reported by \citetalias[][their Table
4]{Dea14}, and according to the errors reported by them, except the angle
$\Omega_b$ which is zero by definition. A set of test runs of the algorithm
systematically gave values of periastrons, mean anomalies, $\Omega_b$, $m_b$,
and $i_b$ at the border of the corresponding intervals. Therefore, we expanded these
initial intervals, in the case of the first five angles to at least one quadrant
(Table~\ref{ic}).
Finally, after investigating the evolution of several
solutions, we found that a difference close to $180^\circ$ between
the arguments of the periastrons of the planets was instrumental to ensure the
long-term stability of the system. Therefore, as an additional constraint, we
imposed that the {initial} values of those arguments differ in
$180^\circ$.

\begin{table}
 \centering
 \caption{Intervals (min, max) into which the elements of the planets and the
  mass of the star were chosen for the first run. ${\bf p}_{\rm f}$: final
  values obtained with our algorithm, at BJD$_{\rm TDB}$ 
 2\,454\,913.8629. 
  }
 \label{ic}
 \begin{tabular}{lrrrr}
  \hline
  Element & min & max & ${\bf p}_{\rm f}$ \\
  \hline
  $m_\star \, [{\rm M}_{\sun}]$ &1.320 &1.470
      &$1.4414540\pm 6\times 10^{-7}$\\
  $a_b \, [{\rm au}]$           & 0.364 &0.377
      &$0.37473469\pm 5\times 10^{-8}$\\
  $e_b$                         & 0.82 &0.91
      &$0.8040\pm 3\times 10^{-4}$\\
  $i_b \, [^\circ]$             & 86.0 &89.1
      &$87.45 \pm 0.09$\\
  $\Omega_b \, [^\circ]$        & 0.0 &0.0
      &0.0\\
  $\omega_b \, [^\circ]$        & 90.0 &270.0
      &$274.43 \pm 0.01$\\
  $M_b \, [^\circ]$             & 60.0 &240.0
      &$125.0\pm 0.2$\\
  $m_b \, [{\rm M_J}]$          & 1.51 &2.80
      &$2.2430\pm 5\times 10^{-4}$\\
  $a_c \, [{\rm au}]$           & 1.650 &1.710
      &$1.70527\pm 7\times 10^{-5}$\\
  $e_c$                         & 0.173 &0.186
      &$0.18715\pm 5\times 10^{-5}$\\
  $i_c \, [^\circ]$             & 84.0 &91.0
      &$87.8\pm 0.9$\\
  $\Omega_c \, [^\circ]$        & $-8.0$ &16.0
      &$9.0\pm 0.1$\\
  $\omega_c \, [^\circ]$        &  &
      &$94.43\pm 0.01$\\
  $M_c \, [^\circ]$             & 180.0 &360.0
      &$227.477\pm 0.007$\\
  $m_c \,[{\rm M_J}]$           & 6.90 &7.70
      &$7.71\pm 0.02$\\
  \hline
 \end{tabular}
\end{table}

Our first run with the extended intervals gave a value of 
$F=358\,397$ for the
best individual, with times measured in days. Computing the mean of the
deviations between the resulting mid-transit times and the observed mid-transit
times yielded about 24 s. We
repeated the run but with angular initial intervals no
greater than one quadrant, choosing the latter according to the final value of
the first run. A fitness $F=478\,590$ was
obtained. Then a third run was 
performed, for which new initial intervals were chosen centered in the values of
the output of the last run, and with widths reduced to 80 per cent of the last
values. This new run gave $F=546\,261$,
equivalent to a mean deviation of about 20
seconds between observed and computed mid-transit times. New attempts gave no
substantial improvement, so we considered the outcome of this third run as our
result.

\begin{figure}
 \includegraphics[width=\columnwidth]{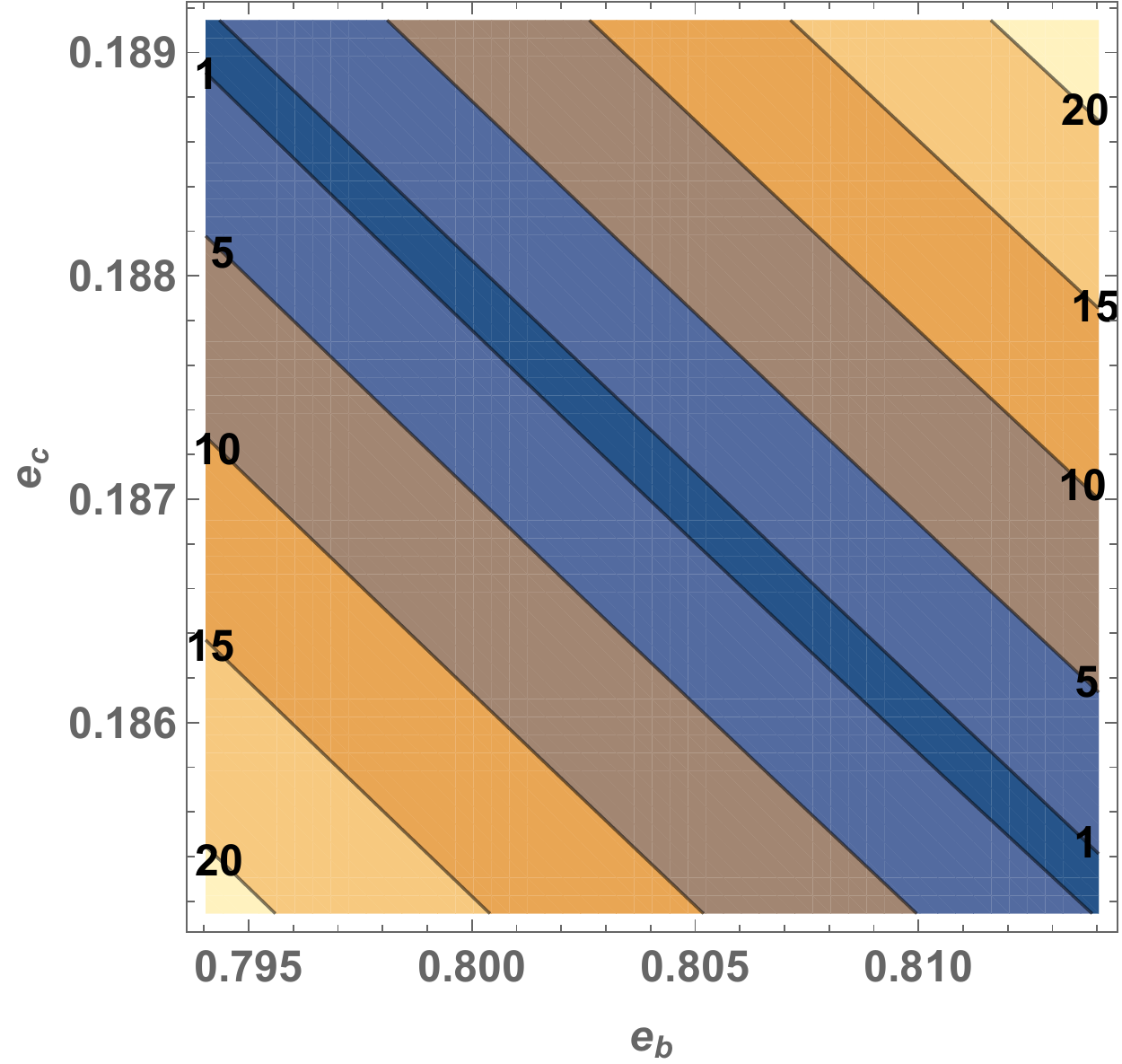}
 \caption{Contours of the root mean square of the differences
between the 
resulting transits and the central values of the observed ones (i.e.,
$1/\sqrt{NF}$), as 
a function of 
the eccentricities of both planets. The contours are labeled with values in
minutes. The narrow central band defines the region into which the
eccentricities should lie in order to obtain one minute or less of error.}
 \label{correl}
\end{figure}

To estimate statistical errors for the parameters, we first
perturbed each 
input time with noise taken from a Gaussian distribution with zero mean and a
dispersion equal to the (maximum) error reported for it in Table 5 of
\citetalias{Dea14}. With this new set of mid-transit times, we repeated the
experiment and registered the output values of the parameters. We repeated this
50 times, and computed the dispersion of the results.
As a check, we inverted the procedure and perturbed the elements
inside the resulting intervals of error, surprisingly obtaining solutions that
went well beyond the errors of the observed
transits. This behavior is expected if, for example, there are correlations
among the elements, because independent perturbations
lead them out of their correlated values, and then the solution
deteriorates. We looked for correlations by taking pairs of elements in
turn, constructing a dense grid of values for each pair 
inside their respective statistical intervals as computed above, and computing
for each point of the grid 
the root mean square of the differences between the resulting 
transits and the central values of the observed ones (i.e., $1/\sqrt{FN}$).
We found several pairs of elements with strong correlation. As an example,
in Fig.~\ref{correl} we show the case $e_b$ versus $e_c$,
where it is seen that the perturbations should lie on a very narrow band
if one wishes to keep around 1 minute of mean error. With the help of these
plots, we computed the true errors of the parameters.
The final values of
the parameters together with their errors
are listed in the fourth 
column of Table~\ref{ic}, corresponding to BJD$_{\rm TDB}$
2\,454\,913.8629. As
a reference, we also integrated the system from this epoch to BJD$_{\rm TDB}$
2\,455\,809.4010, that is, the epoch at which the elements of \citetalias{Dea14} are
defined. The resulting values 
are equal to the ones listed (within errors), with the obvious
exception of the  mean anomalies which gave values of $M_b=69^\circ$ 
and $M_c=345^\circ$.

\begin{figure*}
 \includegraphics[width=\columnwidth]{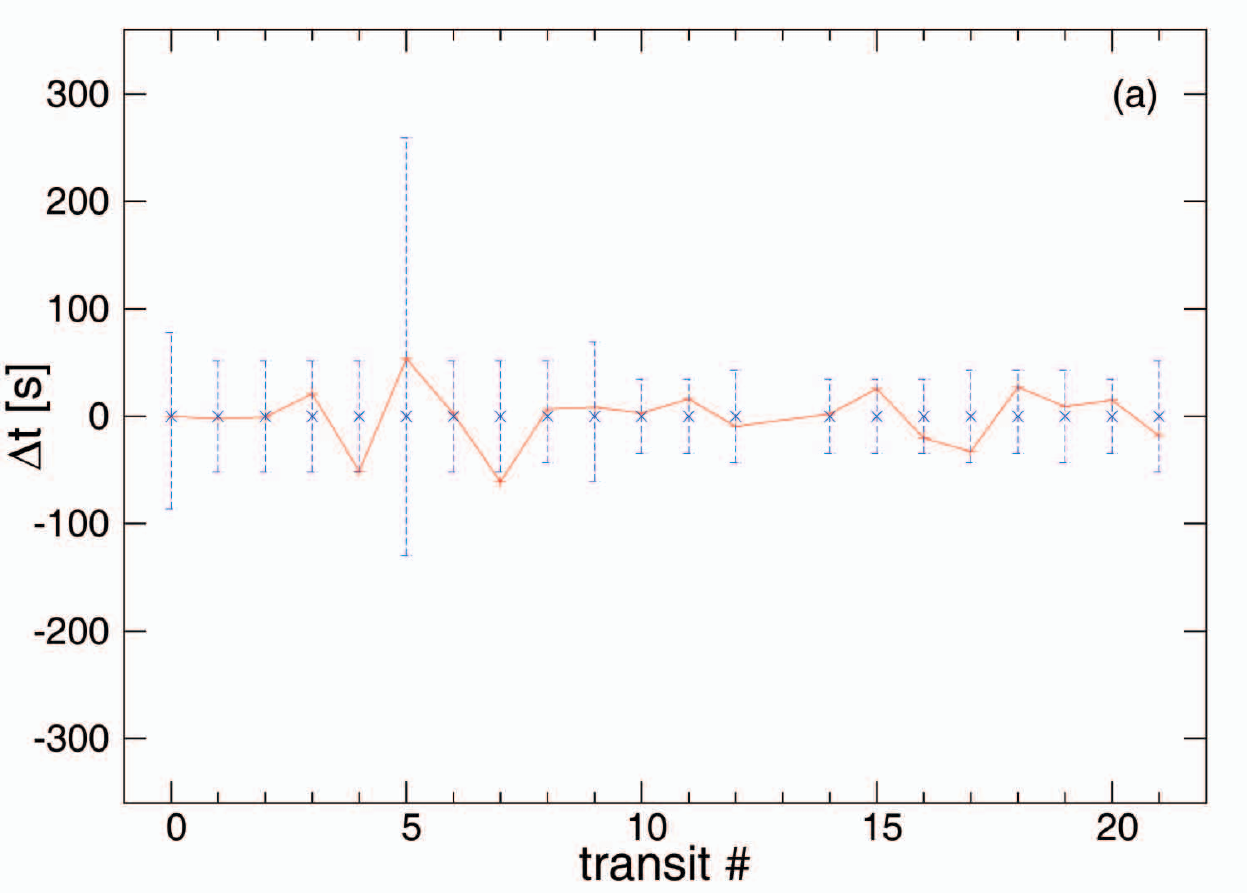}
 \includegraphics[width=\columnwidth]{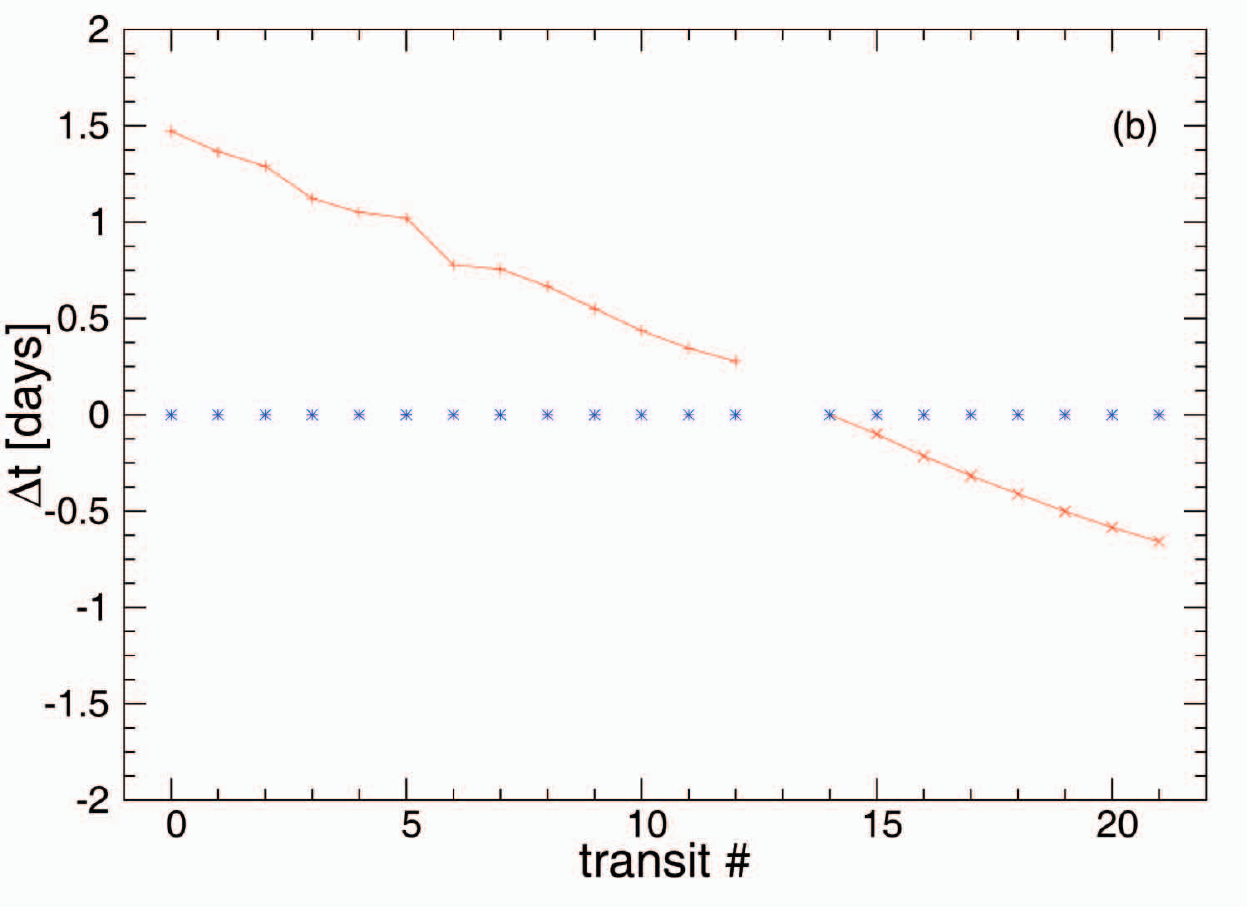}
 \caption{(a) Left: Differences $t_{{\rm com},i}-t_{{\rm obs},i}$ between the
computed transits and the observed ones (plus signs joined with a solid line),
and errors in the observed transits as reported in \protect\citetalias{Dea14}
(crosses with error bars). (b) Right: As in (a) but with the parameters taken
from \protect\citetalias{Dea14}. We note the different scales on the plots.} 
 \label{deltat}
\end{figure*}

Figure~\ref{deltat}a shows the resulting differences $t_{{\rm com},i}-t_{{\rm
obs},i}$ between the computed transits and the observed ones, compared with the
errors in the observed transits reported in \citetalias{Dea14}. The first
transit has a null error by construction. In order to compare the quality of our
solution against the previous one of \citetalias{Dea14}, we also computed the
transits resulting from integrating the orbits that correspond to the central
values of their Table 4. Since they reported that their elements are Jacobian,
and we are not aware of a definition of Jacobian elements, we interpreted this as referring to Jacobian Cartesian coordinates. Thus, we first converted the elements of
\citetalias{Dea14}  to Jacobian Cartesian coordinates
taking the interior planet and the star as the first subsystem
of the ladder; these were in
turn converted to astrocentric Cartesian and integrated. Nevertheless, the
differences between Jacobian and astrocentric Cartesian coordinates are, in this
system, negligible. We also tried an integration by directly converting the  elements of
\citetalias{Dea14} as if they were astrocentric, and no appreciable
differences were found. Another point to take into account is that
these elements are given at BJD$_{\rm TDB}$ 2\,455\,809.4010, so
the integration includes both a backward and a forward period. The forward
integration gave a first transit about 13 days after the initial epoch, which
came as a surprise because that is the interval between the initial epoch and
the {previous} transit. During this process we realized that the coordinate
system of \citetalias{Dea14} is defined with the $x$ axis at the
{descending} node, and the angles $\omega$, $M,$ and $i$ are measured
towards the $-z$ semispace (see Fig. 10 of \citetalias{Dea14}). This implies
that a transit is defined as the passage through $y=0$ from $x>0$ to $x<0$. On
the contrary, in a coordinate system like ours ($x$ axis at the ascending node
and angles measured towards the $+z$ semispace) the transit occurs from $x<0$ to
$x>0$. In terms of orbital elements, the only difference is in the argument of
the periastron: the argument of \citetalias{Dea14} is obtained by adding $\pi$ to
ours (see Appendix \ref{app2}); this should be taken into account if our
elements are to be compared with those of \citetalias{Dea14}. Considering this
difference, we integrated the \citetalias{Dea14} system again, finding a large
drift which causes a difference of the order of 
one day at the extreme points
(Fig.~\ref{deltat}b).
In light of our results regarding the errors of the
elements, we suspected that this drift was probably due to a lack of enough
decimal digits in the solution. We integrated the
solution of \citetalias{Dea14}  again with more digits (Dawson, private communication) and found a
sensible improvement: less than a quarter of a day in the backwards
integration. This supports the need to give all the necessary digits in
Table \ref{ic} in order to reproduce the desired solution.
All the integrations were also reproduced
with the Bulirsh-Stoer method implemented in the {\sc mercury} package
\citep{C99}, with a fixed time step of 1 day and an added routine to reap the
transits.

Table \ref{futuro} provides a list of the transits computed with our solution.
We also include future transits to allow comparison against prospective new
observations.

\begin{table}
 \centering
 \caption{Transits computed by integrating our model.} 
 \label{futuro}
 \begin{tabular}{rrr}
  \hline
  \# & {\rm BJD$_{\rm TDB}$}${}-2\,454\,833$ & date \\
  \hline
  1 & 126.3308& \\
  2 & 196.0608& \\
  3 & 265.7667& \\
  4 & 335.5755& \\
  5 & 405.3159& \\
  6 & 475.0092& \\
  7 & 544.7264& \\
  8 & 614.4553& \\
  9 & 684.1879& \\
 10 & 753.9192& \\
 11 & 823.6435& \\
 12 & 893.3498& \\
 13 & 963.0388& \\
 14 &1032.9388& \\
 15 &1102.6199& \\
 16 &1172.3055& \\
 17 &1242.0117& \\
 18 &1311.7264& \\
 19 &1381.4431& \\
 20 &1451.1570& \\
 21 &1520.8612& \\
$\cdots$ & $\cdots$ & $\cdots$ \\
 50 &3543.0572& 2018-09-14\\
 51 &3612.7392& 2018-11-23\\
 52 &3682.4225& 2019-01-31\\
 53 &3752.3111& 2019-04-11\\
 54 &3822.0053& 2019-06-20\\
 55 &3891.7212& 2019-08-29\\
  \hline
 \end{tabular}
\end{table}

\begin{figure}
 \includegraphics[width=\columnwidth]{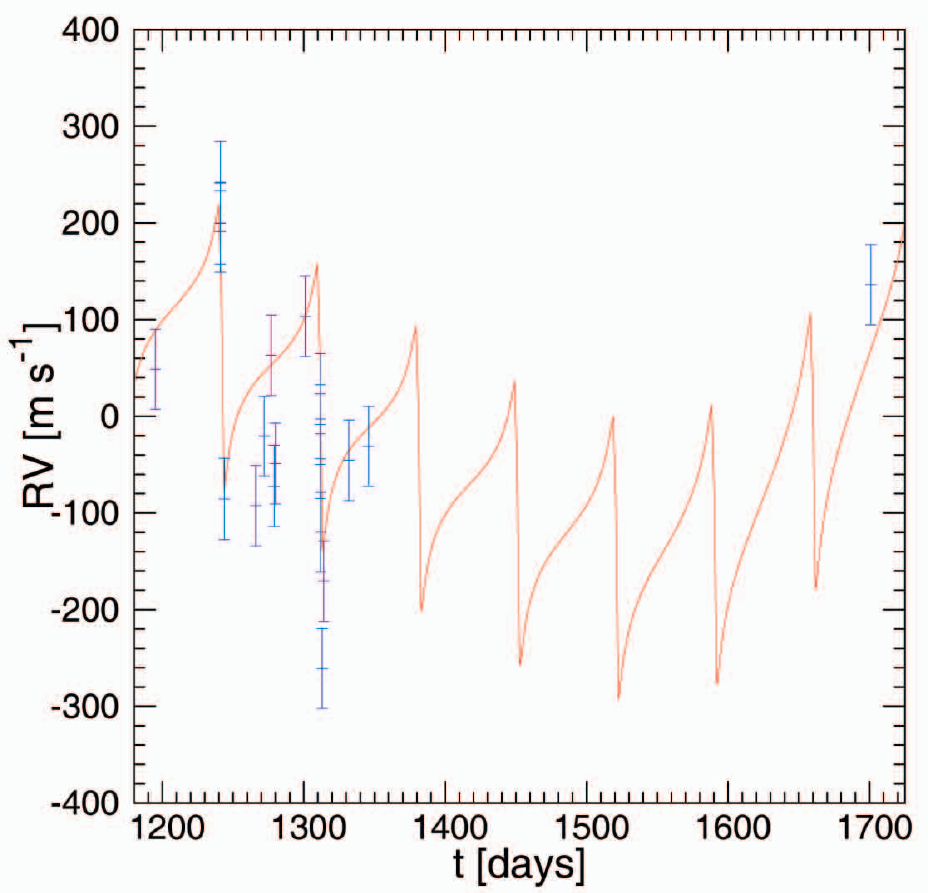}
 \caption{RV of the final system 
  (solid curve), and observed values
  (dots with error bars).} 
 \label{vrad}
\end{figure}

We also computed the radial velocity (RV) of the star with respect to the center
of mass of the system. Since the latter is at the origin of an inertial frame
fixed in space with respect to the observer, the RV is simply the velocity $\dot
z_\star$. This is plotted in Fig.~\ref{vrad} 
(solid line), together with the 
observed values (points with error bars).

\begin{figure}
 \includegraphics[width=\columnwidth]{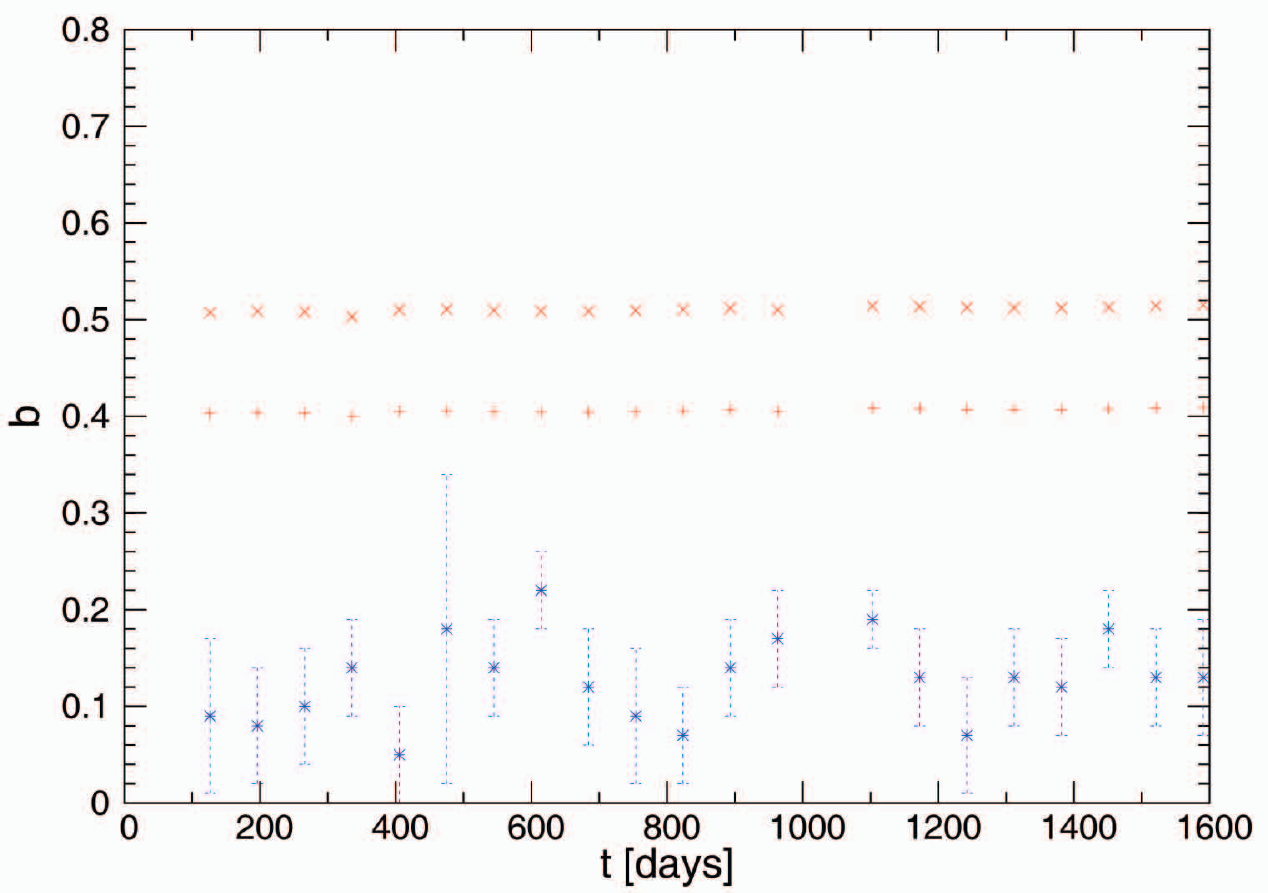}
 \caption{Impact parameter $b$, computed as the fraction of the star radius at
  which the point representing the planet sits when it transits. The upper (red)
  crosses are computed considering a stellar radius $R_\star=1.39\,{\rm
  R}_{\sun}$. The set of pluses correspond to a stellar radius
  $R_\star=1.75\,{\rm R}_{\sun}$. The lower (blue) crosses with error bars are
  the values obtained by \citetalias{Dea14} with the {\sc tap} software
  \citep{GJTDEMA12}.}
 \label{bes}
\end{figure}

Considering that the inclination of the orbit of the TP in our solution is
$87\fdg 4$ (cf. $88\fdg 95$ in
\citetalias{Dea14}), one may wonder whether the 
TP transits at all. We computed the impact parameter $b$ directly from the
dynamical simulation as the difference between the $y$-coordinate of the TP
minus the $y$-coordinate of the star at each transit (both are points
without dimension in the simulation), and compared them to two different
stellar radii: 1.75 R$_{\sun}$, that is, the final value reported in
\citetalias{Dea14}, and 1.39 R$_{\sun}$, the minimum value found by those
authors among the different fits. Figure \ref{bes} shows that the exoplanet indeed
transits in spite of the low inclination of its orbit. We also computed the
duration of the transits, using the formulas of \cite{W11}, taking into account
that a different coordinate system is used in that work, namely our $\omega$ is
Winn's $\omega-\pi$ (see Appendix \ref{app2}). With
$R_\star=1.75\,{\rm R}_{\sun}$, we obtained a total duration of 3.76 hours; with
$R_\star=1.39\,{\rm R}_{\sun}$, the eclipse lasts 2.84 hours, very close
to the reported value in \cite{Dea12} of 2.92 hours.

\subsection{Advantages of using mid-transit times instead of transit timing variations}

The TTV signal is the result of computing the differences between the observed
transit times ($O$) and a linear ephemeris value which corresponds to a periodic
orbit ($C$) for each transit epoch. In this way it is expected that  the deviation of the transits of a planet from a Keplerian
orbit  can be visualized in
an $O-C$ plot. However, the actual linear ephemeris is unknown, and therefore it is estimated from
the mid-transit times themselves, usually by a least-squares fit. Unless the
deviations are evenly distributed above and below the real ephemeris (in a
least-squares sense), this procedure does not guarantee that the computed
linear fit will coincide with the Keplerian
period.

For Kepler-419b, for example, the 
mean period between transists
-- obtained from the linear fit of
the $O-C$
data -- is 69.7546 days \citepalias{Dea14}. However, computing the Keplerian
motion from our solution yields
\begin{equation}
P_b=2\pi\sqrt{\frac{a_b^3}{G m_\star}}=
69.7844 \quad{\rm days},
\end{equation}
that is, about 42 minutes of difference. That is
why we chose to work with the bare 
mid-transit times instead of going through the TTVs.

\section{Dynamics and long-term evolution}

We studied the general orbital dynamics and the long term evolution of the
system using as initial conditions the orbital and physical elements that we
have determined, in order to evaluate its stability. We integrated the orbits
with the Bulirsh-Stoer method implemented in the {\sc Mercury} package
\citep{C99}, with a fixed time step of 0.1 days for a simulated timespan of 200
Myr. 

\begin{figure}
 \includegraphics[width=\columnwidth]{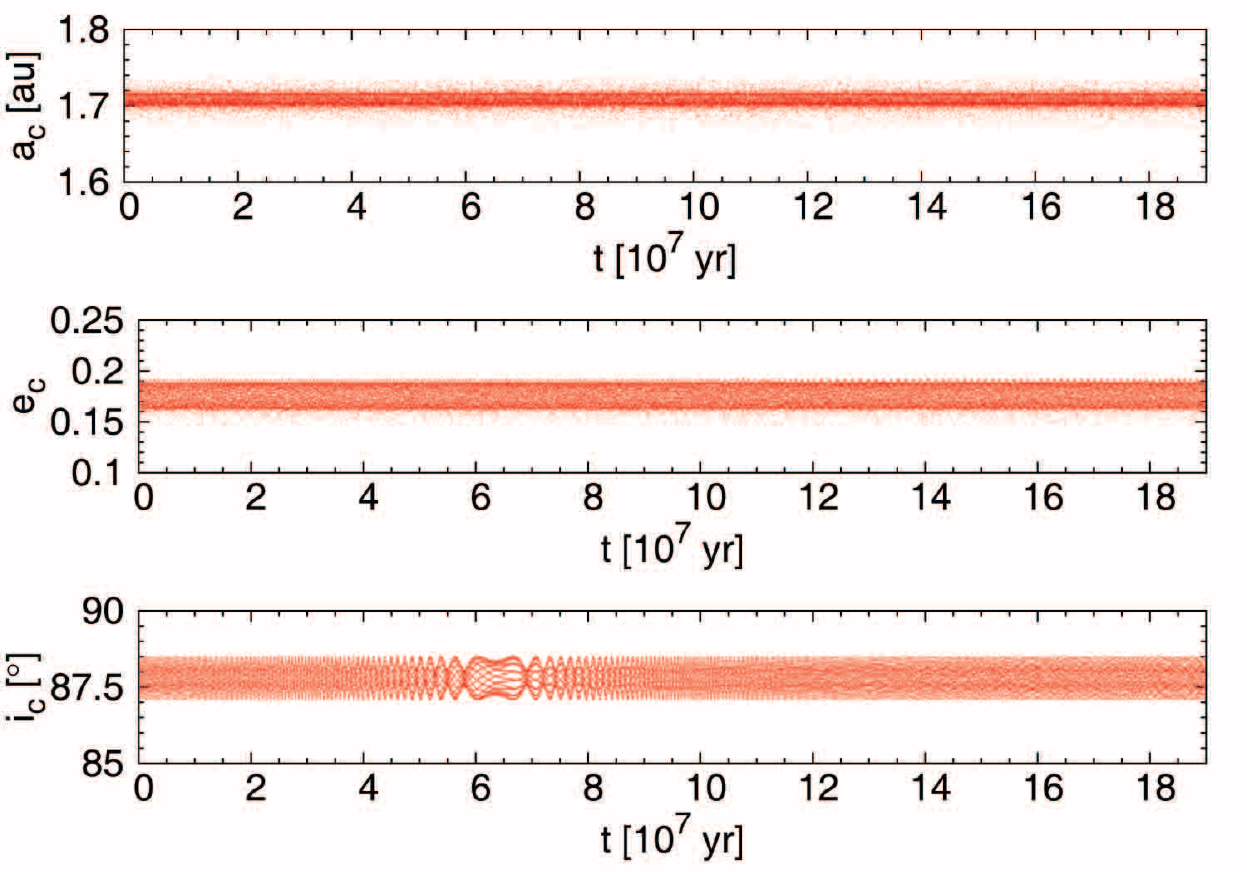}
 \caption{Long-term evolution of the semimajor axis, the eccentricity, and the
  inclination of Kepler-419c using  the elements listed in
  Table \ref{ic} as initial conditions.}
 \label{fig:p2l}
\end{figure}

\begin{figure*}
\sidecaption
 \includegraphics[width=12cm]{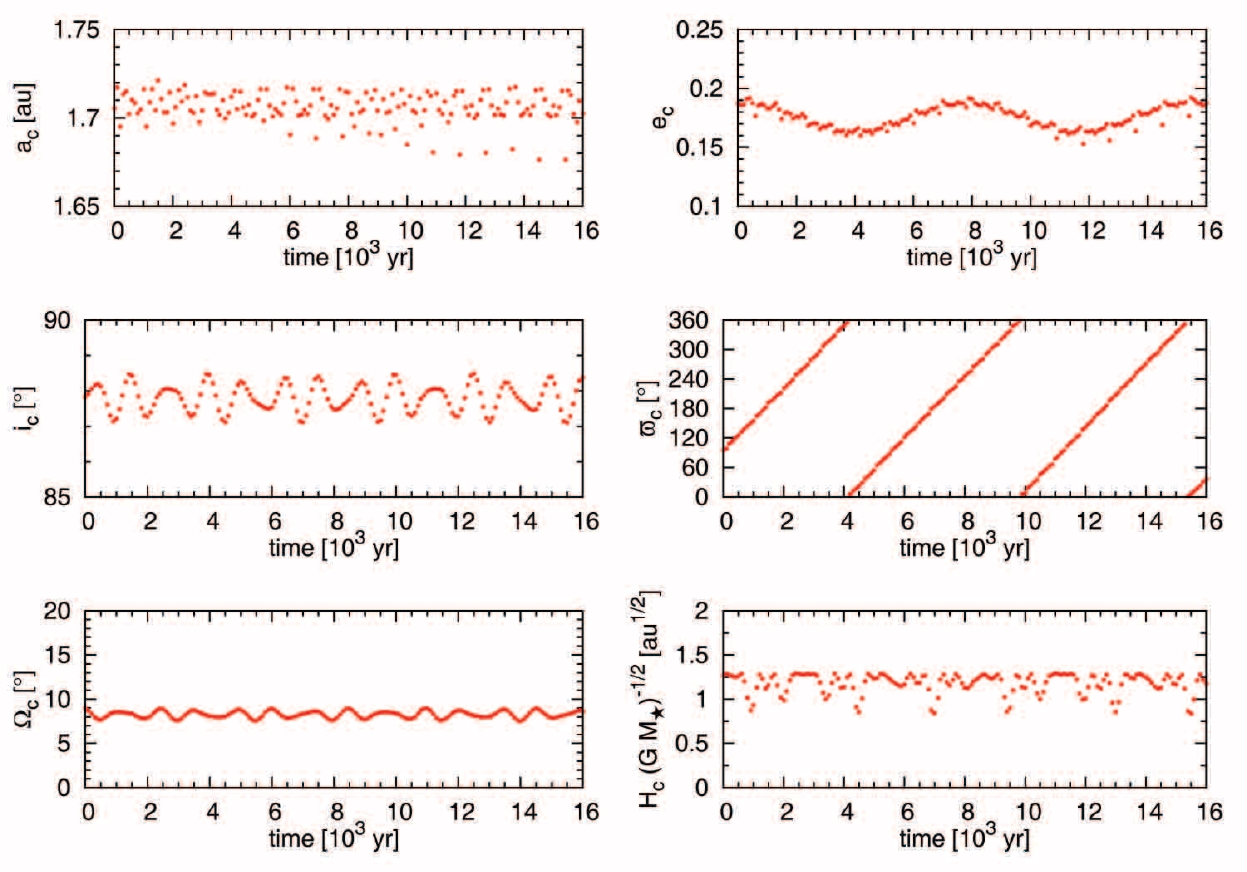}
 \caption{Orbital dynamics of Kepler-419c using the
   elements listed in Table \ref{ic} as initial conditions.}   
 \label{fig:p2}
\end{figure*}

We find that Kepler-419c, the more massive and distant planet, follows a stable
secular behavior (Fig. \ref{fig:p2l}), that is, the mean values of the semimajor
axis, the eccentricity, and the inclination remain remarkably constant. Figure
\ref{fig:p2}, corresponding to the first 16\,000 yr of integration, shows the
short-term evolution of the elements. It is seen that the semimajor axis has a
mean value of $\simeq$ 
1.71  au, whereas the variation about this value has a
small amplitude, of less than 0.05 au. The eccentricity librates about 0.2, the
main component having a period of about 8000 yr and an amplitude of less than
0.05. The inclination also librates about a value of
$\simeq 87\fdg 5$
with a
period of the order of 
3800 yr. The longitude of the periastron rotates with a
period of about 7000 yr. The node librates about a value of $\simeq
13^\circ$ with a period similar to that of the inclination. As expected, the
plot of the Delaunay's variable $H_c=\sqrt{G m_\star a_c(1-e_c^2)} \cos i_c$
shows that it is not conserved, that is, the planet is not in the Kozai
(\citeyear{K62}) resonance. This is further confirmed by the lack of both
coupled oscillations between the eccentricity and the inclination, and libration
of the longitude of the periastron.

\begin{figure}
 \includegraphics[width=\columnwidth]{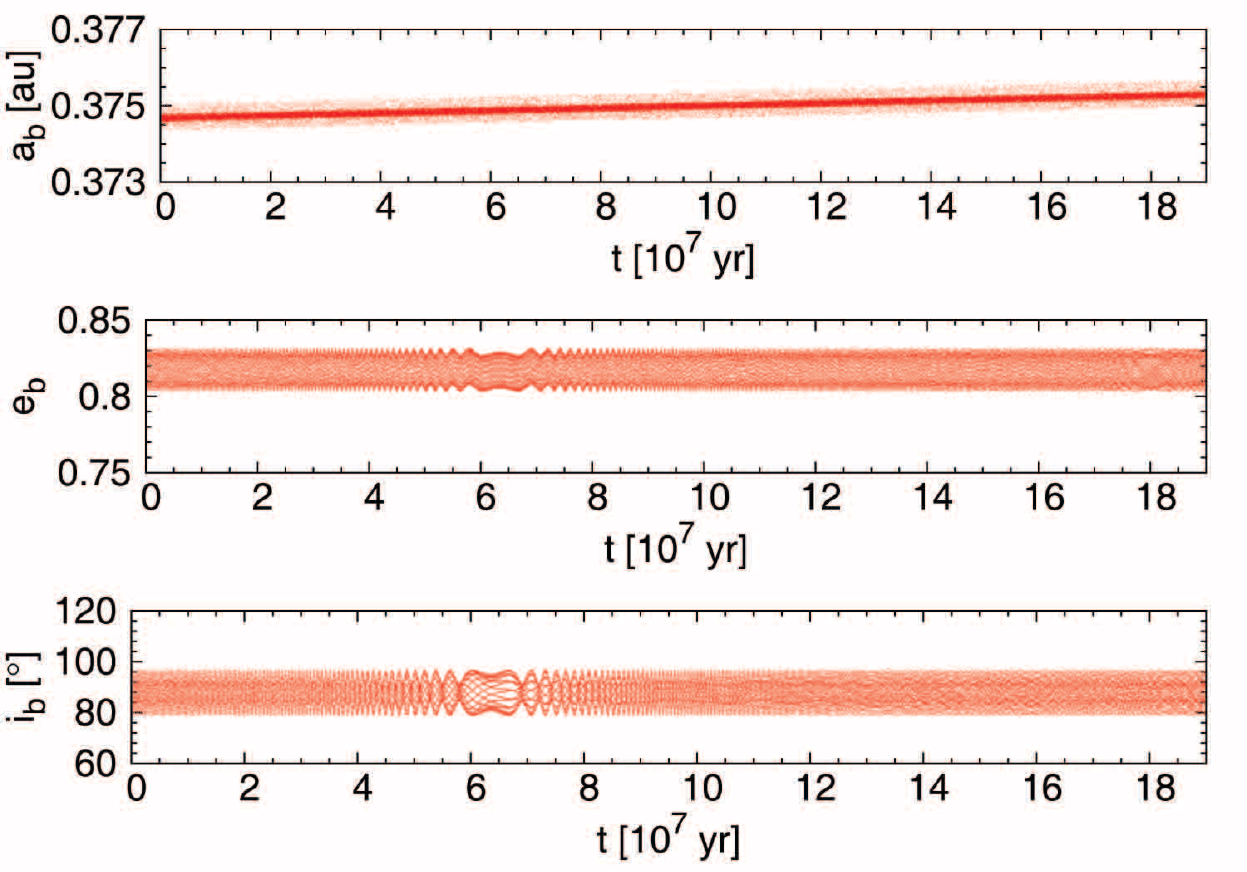}
 \caption{Long-term evolution of the semimajor axis, the eccentricity and the
  inclination of Kepler-419b, using  the elements listed in
  Table \ref{ic} as initial conditions.}
 \label{fig:p1l}
\end{figure}

\begin{figure*}
\sidecaption
 \includegraphics[width=12cm]{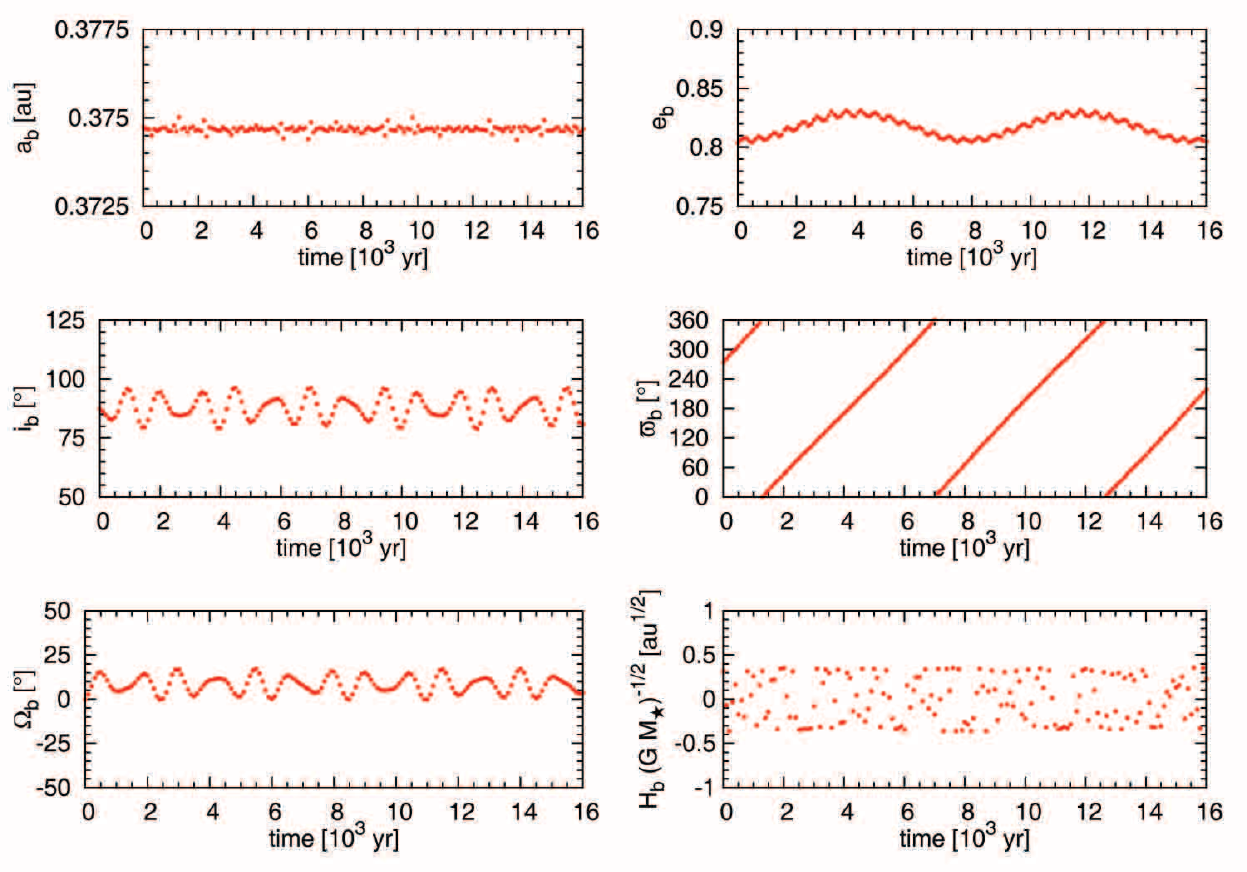}
 \caption{Orbital dynamics of Kepler-419b using the
  elements listed in Table \ref{ic} as initial conditions.}   
 \label{fig:p1}
\end{figure*}

In contrast, the long-term evolution of the inner planet Kepler-419b
(Fig. \ref{fig:p1l}) shows 
a slow variation of its semimajor axis, probably due
to the proximity to a high-order mean-motion resonance, since the
ratio of periods between Kepler-419b and Kepler-419c is
close to
one tenth. On the other hand, the eccentricity and the inclination are as
constant in the long term as in Kepler-419c. Figure \ref{fig:p1} shows
the short-term dynamics. The periods involved
are clearly almost the same as in the other planet. The semimajor axis
maintains a mean value of 
0.375 au, with very small variations around
it. The eccentricity librates harmonically about
0.81, the main
component having a period of about 8000 yr and an amplitude of about
0.03; there is also a clear second-order libration, with a period of
approximately 200 yr. The inclination also librates about a value of
$\simeq
87.5^\circ$, with a period somewhat larger than
2000 yr and a
rather large amplitude of $\simeq 25^\circ$. The longitude of the
periastron rotates with a period of
about 7000 yr.  The node librates about a value of $\simeq 15^\circ$
with a period similar to that of the inclination.
Delaunay's variable $H_b=\sqrt{G m_\star a_b(1-e_b^2)} \cos i_b$ is 
not conserved 
and, in general, the eccentricity and inclination do not
oscillate in counter-phase and the longitude of the periastron
does not librate, except for an event at about 60 Myr where there 
is a temporary capture
in the Kozai's resonance.

\begin{figure}
 \includegraphics[width=\columnwidth]{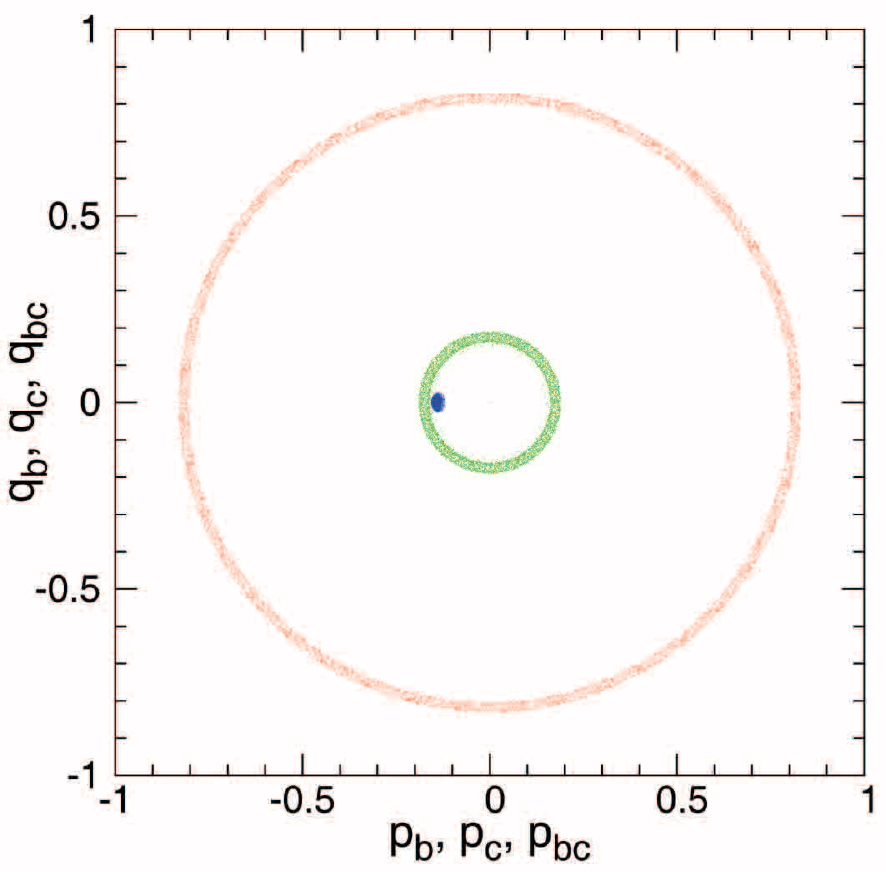}
 \caption{Plots of $p_b$ vs. $q_b$ (red dots), $p_c$ vs. $q_c$ (green dots), and
  $p_{bc}$ vs. $q_{bc}$ (blue dots).}
 \label{fig:p12}
\end{figure}

It is remarkable that both planets are in a $\dot{\varpi_b}=\dot{\varpi_c}$
secular resonance, that is, both longitudes of the periastron rotate at the same
rate. We define the regular elements:
\begin{align}
&p_b = e_b\, \cos \varpi_b,\\
&q_b = e_b\, \sin \varpi_b,\\
&p_c = e_c\, \cos \varpi_c,\\
&q_c = e_c\, \sin \varpi_c,\\
&p_{bc} = e_b\, e_c\, \cos(\varpi_b-\varpi_c),\\
&q_{bc} = e_b\, e_c\, \sin(\varpi_b-\varpi_c).
\end{align}
We plot these elements in Fig. \ref{fig:p12}, where it can be clearly seen that
$\varpi_b-\varpi_c$ librates about a value of $180^\circ$ with a small amplitude.

It is worth noticing that other solutions found by changing, for example, the
seed of the random number generator or the amplitude of the initial intervals,
were sometimes better than the reported one with respect to the fit of the
mid-transit times, but they were unstable. The details of some
of these unstable solutions are given in Table \ref{tab_inest}. After inspecting
several outcomes, it was apparent that an (close) anti-alignment of the arguments
of the periapses all along the evolution of the system was a necessary 
condition for the stability of the system: this ensured that the distance of
closest approach between the planets occured at the maximum possible value, when
one of them was at the periastron and the other one at the apoastron. Although a
more detailed dynamical description of this condition is necessary, it is beyond
the scope of this investigation. In Table \ref{tab_inest} we also report the
outcome of the long-term integration of the elements of the Kepler-419 system as
proposed in \cite{Aetal18} (A18), which turns out to be unstable as well. We note
that in the latter case the longitude of the node of Kepler-419b is $180^\circ$,
due to a different choice of the coordinate system.

\begin{table*}
 \centering
  \caption{Elements of unstable planetary systems $S_i$ with a
  large value of the fitness $F$. The timescale $\tau$ is
  the time for the inner planet to reach a distance larger than 10 au from the
  star, at which point we consider it detached from the system.}
       \label{tab_inest}
        \begin{tabular}{lrrrr}
\hline
Parameters  & $S_1$ & $S_2$ & $S_3$ & A18 \\
  \hline
  $m_\star\, [{\rm M}_\sun]$  & 1.3402 & 1.3362 & 1.3402 & 1.5810 \\
  $a_b \, [{\rm au}]$         & 0.3657 & 0.3654 & 0.3657 & 0.3865 \\
  $e_b$                       & 0.802  & 0.811  & 0.803  & 0.8070 \\
  $i_b \, [^\circ]$           & 87.122 & 87.190 & 87.219 & 87.372 \\
  $\omega_b \, [^\circ]$      & 259.08 & 239.33 & 259.08 & 95.23  \\
  $M_b \, [^\circ]$           & 223.39 & 251.40 & 223.39 & 352.80 \\
  $m_b \, [{\rm M}_{\rm J}]$              & 1.284  & 1.986  & 1.285  & 3.067  \\ 
  $a_c \, [{\rm au}]$         & 1.6625 & 1.6591 & 1.6643 & 1.7520 \\
  $e_c$                       & 0.1877 & 0.1894 & 0.1870 & 0.1797 \\
  $i_c \, [^\circ]$           & 87.236 & 86.470 & 87.298 & 85.720 \\
  $\Omega_c \, [^\circ]$      & $-4.47$& $-4.82$& $-4.70$& 184.77 \\
  $\omega_c \, [^\circ]$      & 82.73  & 65.71  & 82.53  & 276.75 \\
  $M_c \, [^\circ]$           & 236.00 & 237.02 & 236.20 & 248.35 \\
  $m_c \,[{\rm M}_{\rm J}]$               & 7.176  & 7.176  & 7.176  & 8.494  \\
  $\tau \, [10^6\,{\rm yr}] $ &  1.1   & 4.8    & 0.45   & 0.60   \\ 
  $F$                         & 535582 & 190228 & 511254 & --     \\
\hline
 \end{tabular}
 \end{table*}

\begin{figure}
 \includegraphics[width=\columnwidth]{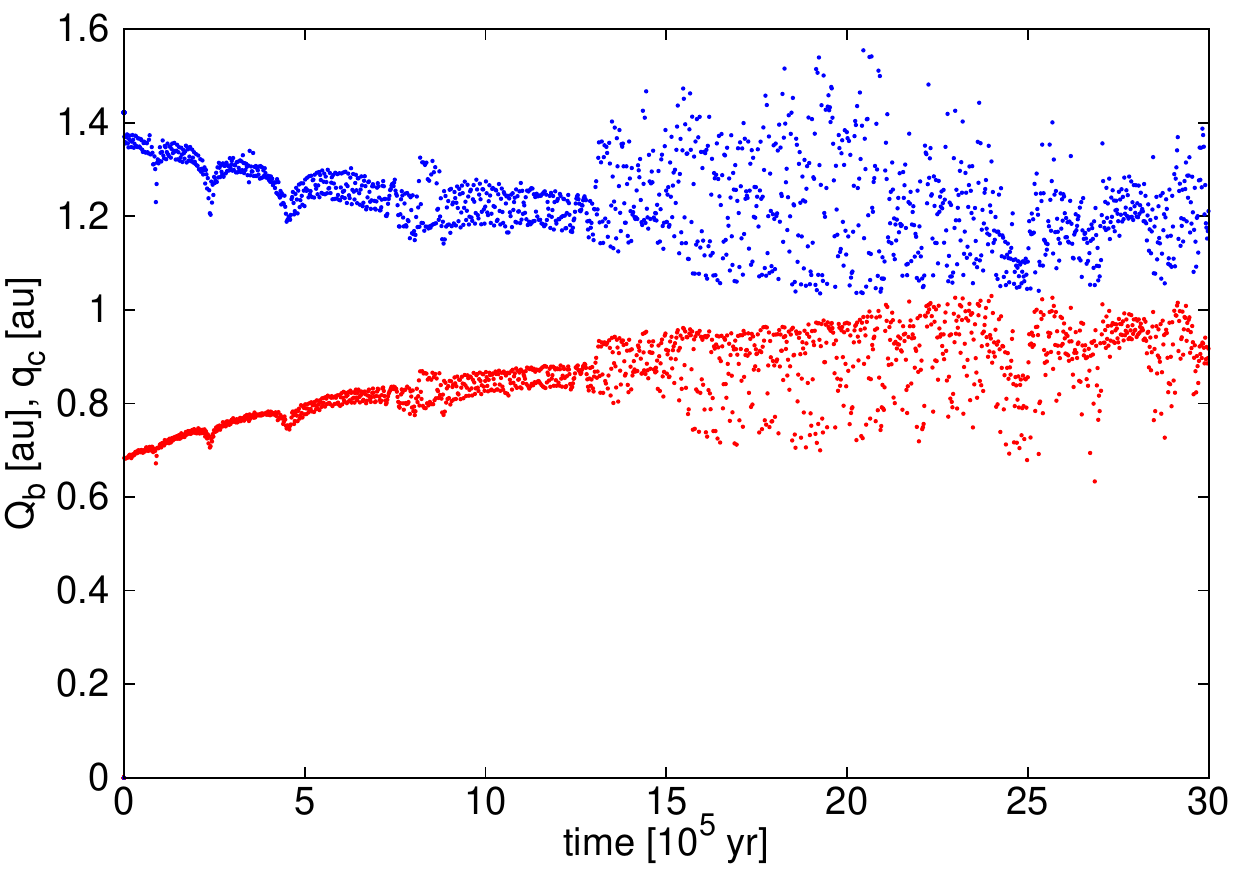}
 \caption{Plot of $Q_b$ (red dots) and $q_c$ (blue dots) as a function of time,
  for an unstable system.}
 \label{uns}
\end{figure}

As an example, Fig.\ref{uns} shows  
the apoastron $Q_b$ of Kepler-419b and the periastron $q_c$ of Kepler-419c as a
function of the time for the unstable system $S_1$;
as can be seen, the
planets suffer a close encounter, the inner one falling on the star as a
consequence of this. Therefore, a solution like this must be discarded in spite of
it yielding a better fit.

\section{Conclusions}

We have solved the inverse problem of the TTVs for the Kepler-419 system,
obtaining remarkable agreement with the observed times of transit, since the
mean of the deviations is of the order of 20 s.

We must note that the methodology that we applied to solve the inverse problem
was able to produce an accurate and stable solution after a reasonable computing
time because we started with a good first guess, as provided by
\citetalias{Dea14}. 
If this information were not available, the extremely
complex landscape of the thirteen-dimensional fitness function would turn any attempt
to find a solution into a daunting if not impossible task. We also note that the
observed transits of Kepler-419b cannot be reproduced using the orbital elements
and masses of the initial solution (i.e., the values of the parameters from
\citetalias{Dea14}) evolved with the equations of the full three-body
gravitational problem integrated with both the Bulirsh-Stoer
subroutine from \citet{PTVF92} and the {\sc Mercury} package
using the option ``BS2'' \citep{C99}.

By finding a solution for the problem, we were able to
compute a set of 
values for the orbital elements of both planets and the mass of the star that
give central transit times within the observational errors.

The dynamics of the Kepler-419 system computed with the initial conditions
listed in Table \ref{ic} is qualitatively similar to the one assuming the
initial elements found by \citetalias{Dea14}. However, the differences are
non-negligible since the mid-transit times differ considerably, meaning that the
validity of one or the other set should requires confirmation through future
observations of the transits of Kepler-419b. It would also be interesting to decipher the quality of the fit of the light curve of the individual transits to models constructed using the orbital and physical parameters determined
in this investigation, which was based solely on the series of transits. 

\begin{acknowledgements}

We want to thank the anonymous referee who pointed out to us
several important aspects of the original draft which greatly improved our
work. We also want to thank Prof. Dr. Juan Carlos Muzzio who gave us useful
advice. 
D.D.C. acknowledges support from the Consejo Nacional de Investigaciones
Cient\'\i ficas y T\'ecnicas de la Rep\'ublica Argentina through grant PIP 0426
and from the Universidad Nacional de La Plata through grant Proyecto 11/G153.
M.D.M. acknowledges the funding by PICT 1144/13 (ANPCyT, MinCyt, Argentina) and
PIP 269/15 (Conicet, Argentina).

\end{acknowledgements}



\bibliographystyle{aa}
\bibliography{dtp}

\begin{appendix} 

\section{H\'enon's step}
\label{henon}

We briefly explain here H\'enon's (\citeyear{H82}) method of landing exactly on
a given plane in only one step of integration.

Consider a dynamical system defined by the $n$ differential equations,
\begin{align}
\label{edif}
&\frac{{\rm d}x_1}{{\rm d}t}=f_1(x_1,\dots,x_n), \nonumber\\
&\vdots\\
&\frac{{\rm d}x_n}{{\rm d}t}=f_n(x_1,\dots,x_n). \nonumber
\end{align}

We want to find the intersections of a solution of Eqs. (\ref{edif}) with an
$(n-1)$-dimensional (hyper-) surface, defined by the equation
\begin{equation}
S(x_1,\dots,x_n)=0.
\end{equation}
To this end, we first define a new variable,
\begin{equation}
x_{n+1}=S(x_1,\dots,x_n),
\end{equation}
and add the corresponding differential equation to the system (\ref{edif}):
\begin{equation}
\frac{{\rm d}x_{n+1}}{{\rm d}t}=f_{n+1}(x_1,\dots,x_n),
\label{new}
\end{equation}
where
\begin{equation}
f_{n+1}=\sum_{i=1}^n f_i \frac{\partial S}{\partial x_i}.
\end{equation}
This allows to define the surface $S$ as the (hyper-)plane
\begin{equation}
x_{n+1}=0.
\end{equation}
Now, we want to turn $x_{n+1}$ into the independent variable: if this is done,
all we have to do to land exactly on $S=0$ is to integrate the new equations
from whatever value $x_{n+1}$ has to 0. The transformation can be easily
achieved by dividing the Eqs. (\ref{edif}) by the new Eq. (\ref{new}), and
inverting the latter:
\begin{align}
&\frac{{\rm d}x_1}{{\rm d}x_{n+1}}=\frac{f_1}{f_{n+1}}, \nonumber\\
\label{edif2}
&\vdots \nonumber\\
&\frac{{\rm d}x_n}{{\rm d}x_{n+1}}=\frac{f_n}{f_{n+1}}, \\
&\frac{{\rm d}t}{{\rm d}x_{n+1}}=\frac{1}{f_{n+1}}. \nonumber
\end{align}
The variable $t$ has now become one of the dependent variables, and $x_{n+1}$
the independent one.

In practice, one integrates until a change of sign is detected for $S$. Then,
one shifts to the system (\ref{edif2}), and integrates it one step from the last
computed point, taking
\begin{equation}
\Delta x_{n+1}=-S
\end{equation}
as the integration step. After that, one reverts to the system (\ref{edif}) to
continue the integration.

The only error involved is the integration error for the system (\ref{edif2}),
which is usually of the same order as the integration error for one step of the
system (\ref{edif}).

\section{Coordinate systems}
\label{app2}

We briefly describe here the differences between the three coordinate systems
mentioned in the text, due to the fact that these differences are somewhat
subtle and not easy to recognize.

\subsection{The standard celestial mechanical system of reference}

\begin{figure}
 \includegraphics[width=\columnwidth]{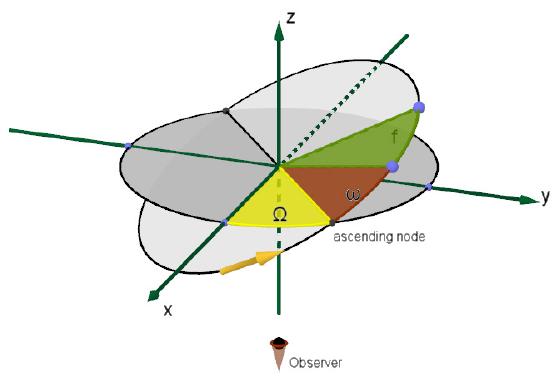}
 \caption{The standard system of reference.}
 \label{murray}
\end{figure}

\cite{MD99} show in their Fig. 2.13 the standard coordinate system used in
Celestial Mechanics (hereafter the standard system); Figure \ref{murray}
reproduces it.

The relevant definitions, taken from \cite{MD99}, are: a) the $x-y$ plane
defines the \emph{reference plane}, and the $+x$ semiaxis defines the
\emph{reference line} of the system; b) the point where the orbit crosses the
reference plane moving from below to above ($-z$ to $z$) is the \emph{ascending
node}; c) the angle between the reference line and the radius vector to the
ascending node is the \emph{longitude of the ascending node}, $\Omega$; d) the
angle between the radius vector to the ascending node and that to the pericentre
of the orbit is the \emph{argument of pericentre}, $\omega$; e) the angle
between the radius vector to the pericentre and that to the planet is the
\emph{true anomaly}, $f$. The planet orbits in the sense of increasing $\omega$
(or $f$).

In the present work, we use this standard coordinate system. We put the
reference plane in the sky (a standard choice) and the observer in Earth at the
$-z$ semiaxis. With this, a transit is defined as the point on the orbit where
$\omega + f=3\pi/2$. Alternatively, the transit may also be defined as the point
where $x=0$ and $\dot x>0$, or as the point where $x=0$ and $z<0$.

\subsection{The Winn's system of reference}

\begin{figure}
 \includegraphics[width=\columnwidth]{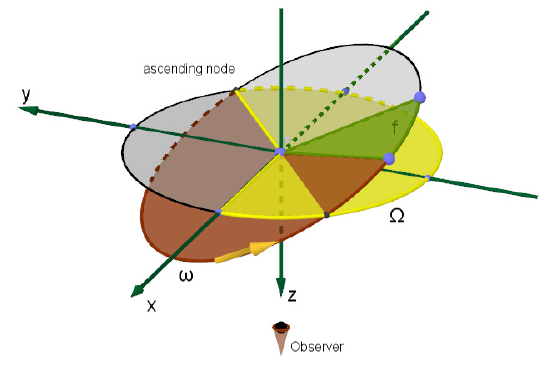}
 \caption{Winn's system of reference.}
 \label{winn}
\end{figure}

\cite{MC10} describe Keplerian orbits using the standard system. In the same
book, \cite{W11} says that ``As in the chapter by Murray and Correia, we choose
a coordinate system centered on the star, with the sky in the $x-y$ plane and
the $+z$ axis pointing at the observer". But then Winn defines the $+x$ semiaxis
pointing to the \emph{descending} node, giving $\Omega=180^\circ$ when the line
of the nodes coincide with the $+x$ axis. It is worth noting that the
descending node in this system is the same point as the ascending node in the
standard system, because the $z$ axis is inverted. Besides, Winn defines the
conjunctions as the condition $x=0$, which gives him $f+\omega=\pi/2$ for the
transit; this means that the argument of pericentre is measured as in the
standard system from the nodal line, that is, from the ascending node. The
transit can also be defined in this system as the point where $x=0$ and $\dot
x>0$ (as in the standard system), or where $x=0$ and $z>0$ (contrary to the
standard system). Another difference is that a prograde orbit in the standard
system (with respect to the $+z$ axis) is a retrograde orbit in this system.
Maintaining the same physical situation than in Fig. \ref{murray}, the elements
of the Winn system are shown in Fig. \ref{winn}.

In terms of values of angles (subindex W stands for Winn, subindex S for
standard), we have $\Omega_{\rm W}=\Omega_{\rm S}+\pi$, and $\omega_{\rm
W}=\omega_{\rm S}+\pi$.

\subsection{The Dawson et al.'s system of reference}

\begin{figure}
 \includegraphics[width=\columnwidth]{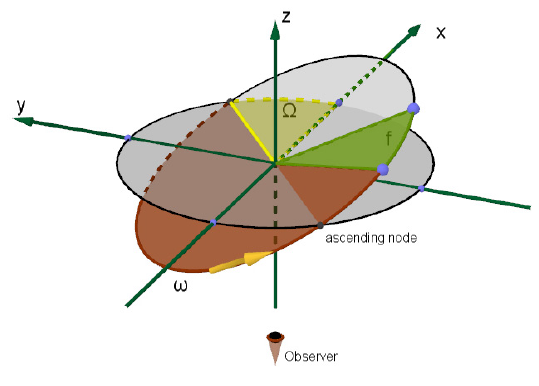}
 \caption{System of reference used in \citetalias{Dea14}.}
 \label{dawson}
\end{figure}

\citetalias{Dea14} used a third system of reference. Their axes have the same
orientation as in the standard system ($x-y$ plane on the sky, $+z$ axis
pointing away from the observer), and also the reference line is on the $+x$
axis. However the argument of the pericentre is measured from the nodal line
towards the $-z$ semispace (see their Fig. 10). Therefore, the nodal
line is in reality the anti-nodal line, and the $+x$ axis correspond to the
{descending node}, as in Winn, though the $+z$ axis is similar to that in the
standard system. Figure \ref{dawson} shows this third system.

In the standard and Winn systems, the line of nodes (ascending node) is the
origin from which  $\omega$ is measured. In the Dawson et al. system, this
origin is the anti-nodal line. Therefore the two classes of systems are defined in an
essentially different way. The same can be said about the definition of
$\Omega$: in the standard and Winn systems, this angle is defined between the
$+x$ axis and the nodal (ascending) line; in the Dawson et al. system, it is
defined between the $+x$ axis and the anti-nodal (descending) line: Dawson et
al. define $\Omega=0$ at this descending node, contrariwise to the standard
definition. The transits are at $\omega+f=\pi/2$, as in Winn. Also they can be
defined as the points where $x=0$ and $\dot x<0$ (contrary to the standard
system), or where $x=0$ and $z<0$ (as in the standard system).

In terms of values of angles (subindex D standing for Dawson et al.), we have
$\Omega_{\rm D} = \Omega_{\rm S} = \Omega_{\rm W}-\pi$, and $\omega_{\rm
D}=\omega_{\rm W}=\omega_{\rm S}+\pi$.

\end{appendix}

\end{document}